\documentclass[aps,prd,reprint,nofootinbib,superscriptaddress]{revtex4-2}

\usepackage[T1]{fontenc}
\usepackage{lmodern}
\usepackage{amsmath,amssymb,bm}
\usepackage{siunitx}
\usepackage{graphicx}
\usepackage{booktabs}
\usepackage{array}
\usepackage{hyperref}
\hypersetup{hidelinks}
\usepackage{xcolor}
\usepackage{mathtools}

\newcommand{\dd}{\mathrm{d}}
\newcommand{\ii}{\mathrm{i}}
\newcommand{\kb}{k_{\mathrm B}}
\newcommand{\SQL}{\mathrm{SQL}}
\newcommand{\zpf}{\mathrm{zpf}}
\newcommand{\imp}{\mathrm{imp}}
\newcommand{\ba}{\mathrm{ba}}
\newcommand{\thm}{\mathrm{th}}
\newcommand{\env}{\mathrm{env}}
\newcommand{\NN}{\mathrm{NN}}
\newcommand{\scg}{\mathrm{sc}}
\newcommand{\add}{\mathrm{add}}
\newcommand{\eqv}{\mathrm{eq}}
\newcommand{\est}{\mathrm{est}}
\newcommand{\rect}{\mathrm{rect}}
\newcommand{\Hann}{\mathrm{Hann}}
\newcommand{\opt}{\mathrm{opt}}

\begin{document}

\title{Torsion balances as operational probes of semiclassical gravity:\\
Matched-filter bounds, torque-diffusion constraints, and quantum-noise benchmarks}

\author{Jyotirmaya Mohanta}
\affiliation{Graduate Program in Computer Science, University of Tsukuba, Japan}

\author{Yutaka Shikano}
\email{yshikano@cs.tsukuba.ac.jp}
\affiliation{Institute of Systems and Information Engineering, University of Tsukuba, Tsukuba, Ibaraki 305-8573, Japan}
\affiliation{Center for Artificial Intelligence Research, University of Tsukuba, Tsukuba, Ibaraki 305-8577, Japan}
\affiliation{Institute for Quantum Studies, Chapman University, Orange, CA 92866, USA}

\date{\today}

\begin{abstract}
Calibrated torsion-balance spectra can be mapped to bounds on deterministic and stochastic
departures from standard gravitational sourcing rules in the weak-field Newtonian regime. Using
one-sided spectra, we formulate a linear calibrated angle-equivalent output-noise budget and the
finite-time matched-filter/Cram\'er--Rao bound for a known torque template. For stochastic
models, the calibrated standard noise budget---thermal, Newtonian, environmental, imprecision,
and backaction terms---is subtracted from the observed angle spectrum to define a residual
spectrum, which is then converted to an equivalent residual torque spectrum using the calibrated
torsional susceptibility. This residual provides an upper bound on any additional stationary
stochastic torque noise. Only in the Markovian white-noise limit can this frequency-resolved
bound be compressed to a single torque-diffusion coefficient $D_\tau$; colored or non-Markovian
models must be compared with the full residual spectrum. Page--Geilker branch discrimination and
Fedida--Kent mixture-equivalence tests address different physical sourcing questions, but after
their competing predictions are projected onto torsion-balance torque templates they become the
same statistical problem of matched-filter template discrimination. For a representative
room-temperature Cavendish benchmark, the resonant thermal angle ASD is
$1.36\times10^{-4}\,\mathrm{rad}/\sqrt{\mathrm{Hz}}$, whereas the measurement-added SQL is
$3.25\times10^{-12}\,\mathrm{rad}/\sqrt{\mathrm{Hz}}$. For the torsion-balance search of Yan
\emph{et al.}, the reported $0.3\,\mu\mathrm{rad}/\sqrt{\mathrm{Hz}}$ sensitivity at
$2.5\,\mathrm{mHz}$ implies a conservative observed-floor bound
$D_\tau \lesssim 2.4\times10^{-23}\,\mathrm{N^2\,m^2\,s}$ when that public floor is interpreted
as white torque noise. These expressions provide a common interface between calibrated
torsion-balance data, deterministic template tests, and stochastic semiclassical-gravity
searches without claiming a direct test of the full relativistic semiclassical Einstein equation.
\end{abstract}

\maketitle

\section{Introduction}
Torsion balances remain laboratory instruments for measurements of small gravitational torques.
They have long been used in determinations of Newton's constant \cite{GundlachMerkowitz2000,Schlamminger2006,Quinn2013}.
They also provide null tests of the weak equivalence principle and of short-range departures from
the inverse-square law \cite{Wagner2012WEP,Kapner2007,Tan2020}. The Page--Geilker experiment
used a torsion balance to confront branch-dependent semiclassical sourcing in a macroscopic
setting \cite{PageGeilker1981}. More recent proposals and experiments use torsional or related
macroscopic mechanical systems as interfaces between quantum measurement and weak gravity
\cite{Yan2025,BoseRMP2025}.

A torsion-balance analysis of semiclassical gravity must keep three literatures conceptually
separate. Linear quantum measurement theory fixes the imprecision--backaction tradeoff and the
measurement-added standard quantum limit (SQL)
\cite{Caves1981,ClerkRMP2010,AspelmeyerRMP2014,BraginskyKhaliliBook,Kimble2001}. Torsion-balance
metrology fixes the calibrated mechanical response and the ordinary thermal, anelastic, seismic,
Newtonian-noise, and technical backgrounds; a separate preprint analyzes atmospheric Newtonian
noise in torsion-balance determinations of $G$ within a GUM-style framework
\cite{MohantaShikano2026ANN}. Semiclassical and postquantum-gravity models then enter only after
that measurement interface has been specified, through deterministic branch-dependent torques,
state-preparation-dependent sourcing rules, or additional stochastic torque noise
\cite{PageGeilker1981,Giulini2022,Oppenheim2023,Layton2023,FedidaKent2025}.

The present paper focuses on this last interface. Big-$G$ metrology, WEP and inverse-square-law
null tests, and quantum-noise budgeting are not treated as new physical results here; they provide
the calibrated language needed to translate an observed angle spectrum into bounds on torque
templates and residual stochastic torque spectra.

Throughout the paper, a \emph{sourcing rule} means a physical or theoretical map from a source
preparation---for example a mass-density history, a quantum state, a branch label, or a reduced
density operator---to the weak gravitational field and hence to the torque template acting on the
balance. It is not a synonym for an experimental error source. Calibration uncertainties,
GUM-style uncertainty budgets, Newtonian-noise backgrounds, and other technical disturbances enter
separately through the measured spectrum and the residual construction below.

The paper makes four modest contributions. First, it standardizes the measurement interface by
keeping one-sided PSD conventions, transfer-function factors, and angle-to-torque conversion
explicit from the output spectrum to the final constraint. These are standard tools, but their
consistent use prevents ambiguities when different gravitational sourcing models are compared with
torsion-balance data. Second, it applies the standard matched-filter/Cram\'er--Rao formalism to
finite-time torsion-balance torque templates for semiclassical and postquantum-gravity tests.
The data-analysis method is not new; the contribution is the explicit projection of candidate
sourcing rules onto calibrated torque templates. Third, it separates a general residual
equivalent-torque spectrum from the special Markovian white-noise compression into a single
experimental summary number $D_\tau$. For colored or non-Markovian models, the spectrum rather
than $D_\tau$ is the comparison object. Fourth, Page--Geilker branch discrimination and the
mixture-equivalence-principle (MEP) discussion of Fedida and Kent are presented as interpretive
applications of the same template formalism. The two tests address different physical sourcing
questions and are not physically equivalent; their commonality appears only after the competing
predictions are projected onto torsion-balance torque templates.

Related collapse-model and Schr\"odinger--Newton studies also require geometry-dependent force or
torque spectra before comparison with data
\cite{BassiRMP2013,Bassi2017,Carlesso2019,Grossardt2021,Altamura2024,Altamura2025}. The present
paper addresses the semiclassical/postquantum-gravity side of the same experimental interface
\cite{Oppenheim2023,Layton2023,FedidaKent2025}.

The paper is organized as follows. Section~\ref{sec:response} summarizes the torsion-balance
response, fixes the one-sided conventions, and states the linear calibrated output spectrum.
Section~\ref{sec:deterministic} treats deterministic torque templates and the finite-time
matched-filter bound. Section~\ref{sec:diffusion} translates residual spectra into stochastic
torque-noise bounds and, only in the white limit, torque-diffusion constraints.
Section~\ref{sec:benchmarks} provides quantitative scale-setting benchmarks for a representative
Cavendish platform and for the recent torsion-balance search of Yan \emph{et al.} \cite{Yan2025}.
Section~\ref{sec:pagekent} treats Page--Geilker branch tests and proper/improper-mixture MEP
tests as interpretive applications of the torque-template formalism. Section~\ref{sec:discussion} collects
experimental design implications. A generic two-channel variational-readout strategy is
summarized in Appendix~\ref{app:variational}; it is included as an extension of the noise model,
not as the central result.

\section{Torsion-balance response and spectral conventions}
\label{sec:response}

\subsection{Mechanical response and notation}
For small angular displacement $\theta$, a torsion balance is modeled as a linear torsional
oscillator with moment of inertia $I$, torsional spring constant $\kappa$, and damping rate
$\gamma$:
\begin{align}
I\ddot{\theta}(t)+I\gamma\dot{\theta}(t)+\kappa\,\theta(t)
={}&
\tau_x(t)+\tau_{\thm}(t)+\tau_{\NN}(t)
\nonumber\\
&+\tau_{\env}(t)+\tau_{\ba}(t).
\label{eq:eom_time}
\end{align}
Here $\tau_x(t)$ denotes whichever signal torque one wishes to constrain: the usual Newtonian
torque in a Big-$G$ measurement, a Yukawa or WEP-violating template, or a branch-dependent or
stochastic semiclassical contribution. The term $\tau_{\NN}$ denotes Newtonian noise, i.e.
gravity-gradient torque from fluctuating environmental mass distributions. It may be grouped
inside a broader environmental torque $\tau_{\env}$, but we keep it explicit when discussing
low-frequency backgrounds and auxiliary-channel subtraction.
With the Fourier convention
$f(t)=\int_{-\infty}^{\infty}\frac{\dd\omega}{2\pi}f(\omega)e^{-\ii\omega t}$,
the torsional susceptibility is
\begin{equation}
\chi(\omega)=\frac{1}{\kappa-I\omega^2-\ii I\gamma\omega},
\label{eq:susceptibility}
\end{equation}
so that $\theta(\omega)=\chi(\omega)\tau(\omega)$ and $\omega_0=\sqrt{\kappa/I}$,
$Q=\omega_0/\gamma$.

We use one-sided power spectral densities (PSDs) for $\omega>0$ throughout. The corresponding
amplitude spectral density (ASD) is $\sqrt{S(\omega)}$. Table~\ref{tab:notation} summarizes the
quantities used most often in the manuscript. Angles are quoted in radians for readability,
although rad is dimensionless in SI.

\begin{table*}[t]
\caption{Symbols and one-sided conventions used in the main text.}
\label{tab:notation}
\centering
\small
\setlength{\tabcolsep}{4pt}
\renewcommand{\arraystretch}{1.12}
\begin{ruledtabular}
\begin{tabular*}{\textwidth}{@{\extracolsep{\fill}}lll}
Symbol & Meaning & Units \\
\colrule
\parbox[t]{0.17\textwidth}{$\chi(\omega)$} &
\parbox[t]{0.57\textwidth}{calibrated torsional susceptibility} &
\parbox[t]{0.17\textwidth}{$\mathrm{rad/(N\,m)}$} \\
\parbox[t]{0.17\textwidth}{$S_{\theta\theta}(\omega)$} &
\parbox[t]{0.57\textwidth}{angle PSD} &
\parbox[t]{0.17\textwidth}{$\mathrm{rad^2/Hz}$} \\
\parbox[t]{0.17\textwidth}{$S_{\tau\tau}(\omega)$} &
\parbox[t]{0.57\textwidth}{torque PSD} &
\parbox[t]{0.17\textwidth}{$\mathrm{N^2\,m^2/Hz}$} \\
\parbox[t]{0.17\textwidth}{$\sqrt{S_{\theta\theta}}$} &
\parbox[t]{0.57\textwidth}{angle ASD} &
\parbox[t]{0.17\textwidth}{$\mathrm{rad}/\sqrt{\mathrm{Hz}}$} \\
\parbox[t]{0.17\textwidth}{$\sqrt{S_{\tau\tau}}$} &
\parbox[t]{0.57\textwidth}{torque ASD} &
\parbox[t]{0.17\textwidth}{$\mathrm{N\,m}/\sqrt{\mathrm{Hz}}$} \\
\parbox[t]{0.17\textwidth}{$D_\tau$} &
\parbox[t]{0.57\textwidth}{stochastic-process white torque-diffusion coefficient, defined by
$S_{\tau\tau}=4D_\tau$} &
\parbox[t]{0.17\textwidth}{$\mathrm{N^2\,m^2\,s}$} \\
\parbox[t]{0.17\textwidth}{$B_w$} &
\parbox[t]{0.57\textwidth}{effective noise bandwidth of the analysis window} &
\parbox[t]{0.17\textwidth}{$\mathrm{Hz}$} \\
\end{tabular*}
\end{ruledtabular}
\end{table*}

For any real stationary process $x(t)$ we use the one-sided/symmetrized relation
$S_{xx}^{\mathrm{1s}}(\omega)=2S_{xx}^{\mathrm{sym}}(\omega)$ for $\omega>0$. In particular,
$S_{\theta\theta}^{\SQL,\mathrm{1s}}=2\hbar|\chi|$ corresponds to
$S_{\theta\theta}^{\SQL,\mathrm{sym}}=\hbar|\chi|$. We reserve $\chi(\omega)$ for the calibrated complex susceptibility of the torsional mode
under study and $|\chi(\omega)|$ for its magnitude. When a specific structural-loss model is
invoked we write $\chi_\phi(\omega)$ explicitly.

For later quantum benchmarks we also introduce the torsional Hamiltonian
\begin{equation}
\hat H=\frac{\hat L^2}{2I}+\frac{\kappa}{2}\hat\theta^2,
\qquad
[\hat\theta,\hat L]=\ii\hbar,
\label{eq:hamiltonian}
\end{equation}
with zero-point angle
\begin{equation}
\theta_{\zpf}=\sqrt{\frac{\hbar}{2I\omega_0}}.
\label{eq:zpf}
\end{equation}

\subsection{Operational weak-field sourcing and torque projection}
The word ``semiclassical'' is used here in an operational weak-field sense. The relativistic
semiclassical Einstein equation is the motivation, but a laboratory torsion balance in the
slow-motion, weak-field regime is described by the Newtonian projection of a candidate sourcing
rule. For a rule $x$ we write schematically
\begin{equation}
\nabla^2\Phi_x(\mathbf r,t)=4\pi G\rho_x^{\mathrm{eff}}(\mathbf r,t),
\label{eq:newtonian_sourcing}
\end{equation}
where $\rho_x^{\mathrm{eff}}$ may be an ordinary classical density, an expectation value, a
branch-conditioned density, or another model-dependent effective source. The balance does not
measure $\Phi_x$ directly. It measures the torque obtained by projecting the weak field onto the
probe mass distribution,
\begin{equation}
\tau_x(t)=\int \dd^3r\,\rho_{\mathrm{probe}}(\mathbf r)
\left[\mathbf r\times\bigl(-\nabla\Phi_x(\mathbf r,t)\bigr)\right]_z,
\label{eq:torque_projection}
\end{equation}
and the observed angle is the calibrated response $\theta_x(\omega)=\chi(\omega)\tau_x(\omega)$.
This is the sense in which torsion-balance data probe semiclassical or postquantum sourcing
rules in this paper: they constrain their observable torque projections, not the full
relativistic field equation itself.

We also distinguish environmental Newtonian noise from a formal fluctuation of Newton's constant.
A replacement $G\to G+\delta G$ can be useful as a theoretical scale-setting exercise in
Schr\"odinger--Newton-type models, but it is not the model of Newtonian noise used below.
Environmental Newtonian noise is generated by fluctuating mass distributions and appears as an
additive torque, a correlated auxiliary-channel contribution, or a multiplicative gravity-gradient
perturbation. If the torsional degree of freedom is quantized, any such scale-setting comparison
should be made with the angular Hamiltonian in Eq.~(\ref{eq:hamiltonian}), rather than with a
free center-of-mass Hamiltonian $\hat P^2/(2M)$.

\subsection{Linear calibrated output spectrum and cross correlations}
We write the measured, angle-equivalent readout as
\begin{equation}
y(\omega)=\theta(\omega)+n_\theta(\omega),
\label{eq:readout}
\end{equation}
where $n_\theta\equiv\theta_{\imp}$ is the angle-referred imprecision noise. Let
$a,b\in\{\thm,\NN,\env,\ba\}$ label thermal torque, Newtonian-noise torque, other
environmental torque, and measurement backaction torque. The most general stationary
one-sided output PSD at the linear level is
\begin{align}
S_{yy}(\omega)
={}& S_{n_\theta n_\theta}(\omega)
+|\chi(\omega)|^2\sum_{a,b} S_{\tau_a\tau_b}(\omega)
\nonumber\\
&+2\,\mathrm{Re}\!\left[
\chi(\omega)\sum_a S_{n_\theta\tau_a}(\omega)
\right].
\label{eq:Stot_general}
\end{align}
This equation is the bookkeeping answer to where Newtonian noise enters: as an additive
gravity-gradient torque $\tau_{\NN}$, and, if it is correlated with some other channel, through
the corresponding cross spectrum. For example, a direct correlation between a gravity-gradient
background and the angle readout would appear as
$2\mathrm{Re}[\chi S_{n_\theta\tau_{\NN}}]$, while correlations between two torque channels
appear inside the double sum.

For the main formulas we use the standard diagonal approximation: thermal noise, Newtonian noise,
other environmental torque noise, and optical backaction are treated as mutually independent,
except for the deliberately engineered or measured imprecision--backaction correlation. Then
Eq.~(\ref{eq:Stot_general}) reduces to
\begin{align}
S_{yy}(\omega)
={}&
S_{\theta\theta}^{\imp}(\omega)
+|\chi(\omega)|^2
\Bigl[S_{\tau\tau}^{\thm}(\omega)
+S_{\tau\tau}^{\NN}(\omega)
\Bigr]
\nonumber\\
&+|\chi(\omega)|^2
\Bigl[S_{\tau\tau}^{\env}(\omega)
+S_{\tau\tau}^{\ba}(\omega)
\Bigr]
\nonumber\\
&+2\,\mathrm{Re}\!\left[
\chi(\omega)S_{n_\theta\tau_{\ba}}(\omega)
\right].
\label{eq:Stot}
\end{align}
Equation~(\ref{eq:Stot}) is the angle-equivalent noise budget used in the rest of the paper:
\emph{every} additive torque-noise PSD enters the measured angle through the same transfer
function $|\chi|^2$. This is elementary from $\theta=\chi\tau$, but it is the key bookkeeping
identity. When correlations other than $S_{n_\theta\tau_{\ba}}$ are non-negligible, one should
return to Eq.~(\ref{eq:Stot_general}) and calibrate the relevant cross spectra. Appendix~\ref{app:variational}
records the corresponding generic two-channel estimator for the case in which correlations are
exploited rather than merely subtracted.

For a quantum-limited linear detector the one-sided spectra obey
\begin{equation}
S_{\theta\theta}^{\imp}(\omega)S_{\tau\tau}^{\ba}(\omega)
-\left|S_{n_\theta\tau_{\ba}}(\omega)\right|^2
\ge \hbar^2.
\label{eq:quantum_ineq}
\end{equation}
In the uncorrelated case $S_{n_\theta\tau_{\ba}}=0$, minimizing the measurement-added part
\begin{equation}
S_{\theta\theta}^{\add}(\omega)
=
S_{\theta\theta}^{\imp}(\omega)
+
|\chi(\omega)|^2 S_{\tau\tau}^{\ba}(\omega)
\label{eq:Sadd}
\end{equation}
gives the standard quantum limit
\begin{equation}
S_{\theta\theta}^{\SQL}(\omega)=2\hbar|\chi(\omega)|.
\label{eq:SQL}
\end{equation}
Because later sections compare thermal, classical, and measurement-added terms on the same
one-sided plots, we explicitly retain the one-sided factors here rather than leaving them
implicit.

\subsection{Thermal torque noise: viscous and structural damping}
For viscous damping, the fluctuation--dissipation theorem gives a white thermal torque PSD,
\begin{equation}
S_{\tau\tau}^{\thm,\mathrm{visc}}(\omega)=4\kb T\,I\gamma,
\label{eq:viscous}
\end{equation}
and therefore
$S_{\theta\theta}^{\thm,\mathrm{visc}}=|\chi|^2S_{\tau\tau}^{\thm,\mathrm{visc}}$.

Real torsion fibers are often better described by a structural-loss model with a
frequency-independent loss angle $\phi$. For the Fourier convention above we write
$\kappa\rightarrow\kappa(1-\ii\phi)$ and define
\begin{equation}
\chi_\phi(\omega)=\frac{1}{\kappa(1-\ii\phi)-I\omega^2}.
\label{eq:chi_struct}
\end{equation}
The one-sided angle thermal PSD is then
\begin{equation}
S_{\theta\theta}^{\thm,\mathrm{str}}(\omega)
=
\frac{4\kb T}{\omega}\operatorname{Im}\chi_\phi(\omega).
\label{eq:structural}
\end{equation}
This form has the correct dimensions and shows that the effective drive noise
is no longer white. Indeed, the angle spectrum can be re-expressed as an equivalent torque PSD
\begin{equation}
S_{\tau\tau}^{\eqv,\mathrm{str}}(\omega)
\equiv
\frac{S_{\theta\theta}^{\thm,\mathrm{str}}(\omega)}{|\chi_\phi(\omega)|^2}
=
\frac{4\kb T\,\kappa\phi}{\omega},
\label{eq:structural_torque}
\end{equation}
which exhibits the $1/\omega$ scaling directly. Within the constant-$\phi$ response model this
conversion is exact algebraically; the caution concerns the validity range of the model itself,
not the equivalent-torque conversion. This distinction matters below, because a stochastic
semiclassical torque source is compared to the equivalent torque ASD rather than to the raw angle ASD. At the same time, Eqs.~(\ref{eq:structural}) and
(\ref{eq:structural_torque}) should be read as an effective band-limited parameterization:
a strictly frequency-independent loss angle extrapolated to arbitrarily low frequency produces
the familiar divergence of the equivalent drive noise and is not a fully causal description in
the $\omega\to0$ limit. In practice, one should use the constant-$\phi$ model only over the
band where it is independently validated for the suspension in question \cite{Saulson1990}.

\subsection{Additive Newtonian noise, multiplicative gradients, and resonance drift}
The damping rate $\gamma$ in Eq.~(\ref{eq:eom_time}) is an effective mechanical loss parameter:
it may include gas damping, internal fiber loss, anelastic loss, or feedback damping. It should
not be interpreted as a ``gravitational viscosity.'' A static external gravity gradient instead
changes the conservative torsional stiffness, for example
$\kappa\rightarrow\kappa+\kappa_g$, and hence shifts the resonance frequency.
Fluctuating environmental mass distributions can therefore enter in two distinct ways. The first
is the additive Newtonian-noise torque $\tau_{\NN}^{\mathrm{add}}(t)$ already included in
Eq.~(\ref{eq:eom_time}). The second is a multiplicative stiffness fluctuation,
\begin{equation}
I\ddot\theta+I\gamma\dot\theta+
\bigl[\kappa+\delta\kappa_{\NN}(t)\bigr]\theta
=
\tau_x+\tau_{\NN}^{\mathrm{add}}+\cdots .
\label{eq:stochastic_stiffness}
\end{equation}
The factor of one half in the resonance drift follows directly from expanding the instantaneous
frequency,
\begin{align}
\omega_0(t)
&=
\sqrt{\frac{\kappa+\delta\kappa_{\NN}(t)}{I}}
=\omega_0\sqrt{1+\frac{\delta\kappa_{\NN}(t)}{\kappa}}
\nonumber\\
&\simeq
\omega_0\left[1+\frac{1}{2}\frac{\delta\kappa_{\NN}(t)}{\kappa}
-\frac{1}{8}\left(\frac{\delta\kappa_{\NN}(t)}{\kappa}\right)^2+\cdots\right],
\label{eq:omega0_taylor}
\end{align}
so that, for $|\delta\kappa_{\NN}|\ll\kappa$,
\begin{equation}
\frac{\delta\omega_0(t)}{\omega_0}
\simeq
\frac{1}{2}\frac{\delta\kappa_{\NN}(t)}{\kappa}.
\label{eq:resonance_drift}
\end{equation}
For example, a fractional stiffness perturbation $|\delta\kappa_{\NN}|/\kappa=10^{-4}$ produces a
fractional resonance shift $|\delta\omega_0|/\omega_0\simeq5\times10^{-5}$. The perturbative
analysis assumes both $|\delta\kappa_{\NN}|/\kappa\ll1$ and stability,
$\kappa+\delta\kappa_{\NN}>0$. If a transient gradient pushed the system close to
$\delta\kappa_{\NN}=-\kappa$, the linear susceptibility used in the matched filter would no
longer be the correct calibrated response; such intervals should be modeled separately or vetoed,
not interpreted as an unbounded Newtonian-noise contribution.

The same stiffness perturbation also shifts the static angular operating point. For a static
calibration torque $\tau_{\mathrm{static}}$, the instantaneous equilibrium is
\begin{align}
\theta_{\mathrm{eq}}(t)
&=
\frac{\tau_{\mathrm{static}}}{\kappa+\delta\kappa_{\NN}(t)}
\simeq
\theta_0\left[1-\frac{\delta\kappa_{\NN}(t)}{\kappa}\right],
\nonumber\\
\theta_0&\equiv\frac{\tau_{\mathrm{static}}}{\kappa}.
\label{eq:theta_eq_stiffness}
\end{align}
Thus $\Delta\theta_{\mathrm{eq}}/\theta_0\simeq-\delta\kappa_{\NN}/\kappa$: a positive gravity-gradient
stiffness reduces the deflection, while a negative stiffness perturbation increases it. Unlike an
additive torque PSD, this multiplicative term convolves the gradient fluctuation with the
oscillator coordinate in frequency space. In a narrowband analysis it is usually handled as a
calibration uncertainty in $\chi(\omega)$, a slow line-frequency drift, or a reduction of the
coherent matched-filter response. Equivalently, it can be monitored with gravity-gradient,
barometric, hydrological, or tilt channels and propagated into the residual through the same
multichannel calibration used for additive Newtonian noise. The formulas below assume that the
susceptibility entering the analysis is the calibrated response over the relevant integration
interval; if the resonance drifts appreciably, the quoted $S_{yy}$ and the leakage factor in
Eq.~(\ref{eq:narrowband}) must include that drift.

\section{Deterministic torque templates and finite-time matched filtering}
\label{sec:deterministic}

\subsection{General template amplitude}
We parameterize a deterministic signal of interest as
\begin{equation}
\tau_x(t)=g_x\,\Lambda_x(t),
\label{eq:template}
\end{equation}
where $\Lambda_x(t)$ is known and $g_x$ is the unknown amplitude parameter to be constrained.
The usual Newtonian torque in a Big-$G$ measurement is obtained by setting $g_x=G$.
A Yukawa deviation at fixed range $\lambda$ may be written as $g_x=\alpha$ with the torque
template $\Lambda_{\alpha,\lambda}(t)$ determined by the attractor geometry. A Page--Geilker
branch comparison corresponds to $g_x=\pm 1$ and $\Lambda_x(t)=\tau_0(t)$.
The advantage of Eq.~(\ref{eq:template}) is that all of these cases share the same estimator.
The ``incoming gravity signal'' in this language is the calibrated torque waveform produced by the
chosen source preparation and modulation protocol.

For a rotating or periodically translated attractor, the template is usually a discrete Fourier
series,
\begin{equation}
\Lambda_x(t)=\sum_{k\ge 1}\Lambda_k
\cos(k\Omega_{\mathrm{rot}}t+\varphi_k),
\label{eq:rotating_template}
\end{equation}
where $\Omega_{\mathrm{rot}}$ is the mechanical modulation frequency and the nonzero harmonics
are fixed by the source geometry. If the attractor has an $N$-fold symmetry, the first useful
line is often at $\omega_m=N\Omega_{\mathrm{rot}}$, with additional harmonics depending on the
mass pattern and shielding. Thus there is no universal gravitational-signal frequency; the
operating frequencies are the Fourier lines of the source torque template that are intentionally
placed inside the calibrated torsion-balance band.

The angle response to the template is
\begin{equation}
\theta_x(\omega)=g_x\,\chi(\omega)\Lambda_x(\omega).
\label{eq:theta_signal}
\end{equation}
For stationary Gaussian noise with one-sided PSD $S_{yy}$, the Fisher information for $g_x$ is
\begin{equation}
\mathcal I(g_x)
=
4\int_0^\infty\frac{\dd\omega}{2\pi}
\frac{\left|\chi(\omega)\Lambda_x(\omega)\right|^2}{S_{yy}(\omega)},
\label{eq:fisher_general}
\end{equation}
and the Cram\'er--Rao bound is
\begin{equation}
\delta g_x \ge \mathcal I(g_x)^{-1/2}.
\label{eq:CRB_general}
\end{equation}
Equation~(\ref{eq:fisher_general}) is the central deterministic result of the paper.
It maps any calibrated output spectrum to a bound on the amplitude of any known torque waveform.

\subsection{Single-tone modulation and explicit window factors}
Suppose
\begin{equation}
\Lambda_x(t)=\Lambda_0\cos(\omega_m t+\varphi)
\label{eq:harmonic_template}
\end{equation}
is observed for a finite integration time $T_{\mathrm{int}}$.
If the analysis is performed with a window $w(t)$, the relevant order-one factor is the
equivalent noise bandwidth
\begin{equation}
B_w
=
\frac{\int_0^{T_{\mathrm{int}}}\dd t\,w^2(t)}
{\left[\int_0^{T_{\mathrm{int}}}\dd t\,w(t)\right]^2}
=
\frac{\eta_w}{T_{\mathrm{int}}},
\label{eq:Bw_def}
\end{equation}
with $\eta_{\rect}=1$ for a rectangular window and $\eta_{\Hann}=3/2$ for a Hann window.
When the modulation frequency is tracked well enough that leakage is negligible, the amplitude bound becomes
\begin{equation}
\frac{\delta g_x}{|g_x|}
\simeq
\frac{\sqrt{S_{yy}(\omega_m)\,B_w}}
{|g_x|\,|\chi(\omega_m)|\,|\Lambda_0|}.
\label{eq:narrowband}
\end{equation}
The approximate $T_{\mathrm{int}}^{-1/2}$ scaling follows from coherent statistical averaging of a
stationary narrowband process. It should not be interpreted as a thermodynamic reduction of the
entropy of the local oscillator or of the gravitational environment. It also holds only while the
noise floor and the response remain approximately stationary over the analysis window. If slow
Newtonian-noise backgrounds, tilt, thermal drift, or multiplicative stiffness fluctuations change
the mean, the correlation structure, or the resonance frequency during the run, increasing
$T_{\mathrm{int}}$ need not improve the uncertainty with the ideal square-root law.

If the phase is also unknown and must be estimated from two orthogonal quadratures, the
right-hand side acquires the usual factor $\sqrt{2}$. If slow eigenfrequency drift or template
mismatch reduces the coherent response, one should further replace $|\Lambda_0|\to
\mathcal L|\Lambda_0|$ with a leakage factor $\mathcal L\le 1$. A useful definition is the
normalized overlap between the windowed true signal $s(t)$ and the windowed template $h(t)$,
\begin{equation}
\mathcal L=
\frac{\left|\int_0^{T_{\mathrm{int}}}\dd t\,w(t)s(t)h(t)\right|}
{\left[\int_0^{T_{\mathrm{int}}}\dd t\,w^2(t)s^2(t)
\int_0^{T_{\mathrm{int}}}\dd t\,w^2(t)h^2(t)\right]^{1/2}}.
\label{eq:leakage_factor}
\end{equation}
For a rectangular window and a pure frequency offset $\Delta\omega$ this reduces, in the
many-cycle approximation, to $\mathcal L\simeq|\mathrm{sinc}(\Delta\omega T_{\mathrm{int}}/2)|$;
for a Hann or other window it is the corresponding normalized window-transform overlap. In real
millihertz analyses, nonstationary drifts can also raise the local floor $S_{yy}(\omega_m)$
itself through spectral leakage into the analysis band. The conservative procedure is therefore
to estimate $S_{yy}(\omega_m)$ from data processed with the same windowing, segmentation,
line-removal, and drift-tracking steps used in the matched filter.

Equation~(\ref{eq:narrowband}) is the precise finite-time version of the qualitative statement
that one should maximize $|\chi|^2/S_{yy}$ within the trusted analysis band. If the dominant
background is white torque noise, the best band is near resonance. If the dominant background
is drift or structural-loss torque noise, the optimum shifts upward. In all cases the relevant
quantity is the measured spectrum itself rather than a generic appeal to ``working near
resonance.''

\subsection{Operational band and noise-frequency taxonomy}
The analysis is not intrinsically restricted to the ultralow-frequency band. The useful operating
band for a given experiment is the set of frequencies where (i) the signal template has support,
(ii) the susceptibility and auxiliary-channel calibrations are reliable, and (iii) the Fisher
kernel
\begin{equation}
\mathcal K_x(\omega)
\equiv
\frac{|\chi(\omega)\Lambda_x(\omega)|^2}{S_{yy}(\omega)}
\label{eq:fisher_kernel}
\end{equation}
is large. Millihertz operation is natural for many torsion balances because the fundamental
torsional resonances are often in that range and slow source modulation is technically clean, but
the same equations apply to higher-frequency modulation until the torsional response, internal
modes, or readout noise make $\mathcal K_x$ small.

Representative examples of the standard terms are cited at the points where they enter the
model: torsion-balance metrology and structural thermal noise in Refs.~\cite{GundlachMerkowitz2000,Schlamminger2006,Saulson1990},
linear quantum measurement and optomechanical readout in Refs.~\cite{Caves1981,ClerkRMP2010,AspelmeyerRMP2014,Yan2025},
and environmental Newtonian noise in Refs.~\cite{Harms2015,MohantaShikano2026ANN}. Table~\ref{tab:freqclasses}
therefore does not try to assign universal numerical frequency bands. Such bands are apparatus
and site dependent. Instead, it records the diagnostic role of each component in the calibrated
one-sided spectrum.

\begin{table*}[t]
\caption{Diagnostic map of the main one-sided spectral components in a calibrated torsion-balance
analysis. The table is not a universal frequency taxonomy; the relevant band is determined by the
measured local floor $S_{yy}$, the calibrated susceptibility $\chi$, and the torque template
$\Lambda_x$.}
\label{tab:freqclasses}
\centering
\scriptsize
\setlength{\tabcolsep}{3pt}
\renewcommand{\arraystretch}{1.18}
\begin{ruledtabular}
\begin{tabular*}{\textwidth}{@{\extracolsep{\fill}}llll}
Component & Angle-equivalent entry & Common spectral character & Interpretation here \\
\colrule
\parbox[t]{0.17\textwidth}{Thermal suspension noise} &
\parbox[t]{0.25\textwidth}{$|\chi|^2S_{\tau\tau}^{\thm}$; Eq.~(\ref{eq:viscous}) for viscous loss and Eq.~(\ref{eq:structural_torque}) for structural loss.} &
\parbox[t]{0.25\textwidth}{White torque noise for viscous damping; $1/\omega$ equivalent torque noise for constant structural loss.} &
\parbox[t]{0.25\textwidth}{Calibrated standard floor. It must be included in $S_{yy}^{\mathrm{std}}$, not counted as an exotic residual.} \\
\parbox[t]{0.17\textwidth}{Newtonian/environmental gravity gradients} &
\parbox[t]{0.25\textwidth}{$|\chi|^2S_{\tau\tau}^{\NN}$ for additive torque noise; $\delta\kappa_{\NN}(t)$ for multiplicative stiffness drift.} &
\parbox[t]{0.25\textwidth}{Usually colored at low frequency and site dependent; may contain deterministic transients or coherent lines.} &
\parbox[t]{0.25\textwidth}{Standard environmental background unless deliberately used as a deterministic torque template or auxiliary-channel subtraction target.} \\
\parbox[t]{0.17\textwidth}{Readout imprecision} &
\parbox[t]{0.25\textwidth}{$S_{\theta\theta}^{\imp}$ in angle units; $S_{\theta\theta}^{\imp}/|\chi|^2$ in equivalent torque units.} &
\parbox[t]{0.25\textwidth}{Often approximately white over a local readout band; torque-equivalent impact grows where $|\chi|$ is small.} &
\parbox[t]{0.25\textwidth}{Part of the measured local floor and the measurement-added noise budget.} \\
\parbox[t]{0.17\textwidth}{Backaction and calibrated correlations} &
\parbox[t]{0.25\textwidth}{$|\chi|^2S_{\tau\tau}^{\ba}+2\mathrm{Re}[\chi S_{n_\theta\tau_{\ba}}]$, with possible auxiliary cross spectra.} &
\parbox[t]{0.25\textwidth}{Detector dependent; constrained by the imprecision--backaction inequality and altered by variational readout.} &
\parbox[t]{0.25\textwidth}{Measurement-added noise. It can be optimized or subtracted only when the relevant cross spectra are calibrated.} \\
\parbox[t]{0.17\textwidth}{Deterministic gravity template} &
\parbox[t]{0.25\textwidth}{$g_x\chi(\omega)\Lambda_x(\omega)$ in the measured angle channel.} &
\parbox[t]{0.25\textwidth}{Known waveform, rotating-source harmonics, or narrow coherent lines.} &
\parbox[t]{0.25\textwidth}{Estimated with the matched-filter/Cram\'er--Rao statistic; covers Big-$G$, WEP/ISL, Page--Geilker, and MEP templates after projection.} \\
\parbox[t]{0.17\textwidth}{Candidate stochastic gravitational torque} &
\parbox[t]{0.25\textwidth}{$|\chi|^2S_{\tau\tau}^{\scg}(\omega)$ added only after the standard budget is specified.} &
\parbox[t]{0.25\textwidth}{Model dependent: white, Lorentzian, $1/f^n$, OU-type, or other non-Markovian spectra.} &
\parbox[t]{0.25\textwidth}{Constrained as a residual equivalent-torque spectrum; compressible to $D_\tau$ only in the Markovian white-noise limit.} \\
\end{tabular*}
\end{ruledtabular}
\end{table*}

Thus the practical design question is to identify an operational window, not to assume in advance
that the lowest accessible frequencies are always best. If Newtonian noise dominates a
millihertz band, moving the template to a higher harmonic can help; if the response has already
rolled off and imprecision dominates, moving still higher can hurt. This tradeoff is the
torsion-balance analogue of the noise-budget optimization used in precision interferometers: the
classical environmental, thermal, and quantum measurement terms must be separated before the best
science band is obvious.

\subsection{Deterministic null tests: WEP and inverse-square-law searches}
The same template language immediately covers the standard torsion-balance null tests. In a WEP
experiment the signal of interest is a composition-dependent differential torque; in an
inverse-square-law search the signal is the Yukawa template generated by
\begin{equation}
V(r)=-\frac{Gm_1m_2}{r}\Bigl[1+\alpha e^{-r/\lambda}\Bigr].
\label{eq:yukawa}
\end{equation}
The experimental result is therefore a bound on
$g_x$ with the appropriate $\Lambda_x$. Our point is not that WEP or Yukawa searches are new,
but that they furnish classical-source calibration data against which any semiclassical or
postquantum model must reduce to the standard macroscopic limit.

\section{Stochastic semiclassical gravity as residual torque noise}
\label{sec:diffusion}

\subsection{Residual spectra and model-independent bounds}
Consistent classical-quantum gravity frameworks generically predict that the classical sector
must be noisy or diffusive when coupled to quantum matter \cite{Oppenheim2023,Layton2023}.
At torsion-balance level the most economical phenomenology is therefore an additional stationary
torque-noise term with one-sided PSD $S_{\tau\tau}^{\scg}(\omega)$:
\begin{equation}
S_{yy}^{\mathrm{obs}}(\omega)=
S_{yy}^{\mathrm{std}}(\omega)+|\chi(\omega)|^2S_{\tau\tau}^{\scg}(\omega),
\label{eq:Scg_def}
\end{equation}
where $S_{yy}^{\mathrm{std}}$ contains the calibrated standard contributions: thermal suspension
noise, Newtonian and other environmental torque backgrounds, readout imprecision, backaction,
calibration uncertainty, and any validated cross spectra. The candidate semiclassical or
postquantum stochastic term is not identified with these ordinary stationary backgrounds; it is
an additional residual term after the standard budget has been accounted for.

Formally, the residual estimate is
\begin{equation}
\widehat S_{yy}^{\mathrm{res}}(\omega)
\equiv
\widehat S_{yy}^{\mathrm{obs}}(\omega)-\widehat S_{yy}^{\mathrm{std}}(\omega),
\label{eq:residual}
\end{equation}
and a physical excess-noise interpretation gives the pointwise upper bound
\begin{align}
S_{\tau\tau}^{\scg}(\omega)
&\le
\frac{S_{yy}^{\mathrm{res},+}(\omega)}{|\chi(\omega)|^2},
\nonumber\\
S_{yy}^{\mathrm{res},+}
&\equiv
\max\{\widehat S_{yy}^{\mathrm{res}},0\}
\quad\text{or an upper confidence band}.
\label{eq:pointwise_bound}
\end{align}
Equation~(\ref{eq:residual}) is simple algebraically but difficult experimentally. The standard
model must include thermal noise, Newtonian noise, environmental drift, seismic tilt, readout
imprecision, backaction, calibration uncertainty, and other technical disturbances. If any
component is incomplete or over-subtracted, individual frequency bins of
$\widehat S_{yy}^{\mathrm{res}}$ can become negative or falsely positive. In practice one should
quote uncertainty bands, confidence intervals, or band-averaged limits rather than treating the
residual as an exact positive function at every frequency. When a fully subtracted residual is
unavailable, Eq.~(\ref{eq:pointwise_bound}) can still be used conservatively by replacing the
residual upper band with the published total observed floor.

For colored low-frequency searches the main challenge is often the construction of the residual
itself: atmospheric pressure, hydrological motion, seismic displacement, tilt, or local moving
masses can generate broad $1/f^n$, Lorentzian, or otherwise colored Newtonian-noise spectra that
mimic non-Markovian semiclassical torque noise. Without auxiliary environmental channels and
cross-spectral subtraction, such a residual should be interpreted as an upper bound on additional
stochastic torque noise, not as evidence for a new gravitational-noise source.

\subsection{White torque diffusion}
For a white stochastic torque process we define $D_\tau$ by the one-sided relation
\begin{equation}
S_{\tau\tau}^{\scg}(\omega)=4D_\tau.
\label{eq:Dtau_def}
\end{equation}
Here ``diffusion coefficient'' is used in the stochastic-process sense: a white random torque
causes diffusive growth of the angular-momentum variance, as shown in
Appendix~\ref{app:dtau_diffusion}. It is not a spatial diffusion constant and does not mean that
a field diffuses through space or decays in amplitude. This scalar compression applies only in the
Markovian white-noise limit. If a candidate semiclassical, postquantum, or collapse-inspired model
predicts colored or non-Markovian torque noise---for example $1/f$, $1/f^2$, Lorentzian, or
Ornstein--Uhlenbeck-type spectra---the correct comparison is not to a single $D_\tau$ but directly
to the frequency-resolved bound in Eq.~(\ref{eq:pointwise_bound}) or its band-averaged version in
Eq.~(\ref{eq:Dtau_band}). Equations~(\ref{eq:pointwise_bound}) and (\ref{eq:Dtau_def}) then imply
\begin{equation}
D_\tau
\le
\frac{S_{yy}^{\mathrm{res},+}(\omega)}{4|\chi(\omega)|^2}
\qquad
\text{(pointwise bound)}.
\label{eq:Dtau_point}
\end{equation}
A band-averaged version is obtained by weighting each frequency bin by the estimator used in the
actual analysis:
\begin{equation}
D_\tau
\le
\frac{
\int_{\mathcal B}\frac{\dd\omega}{2\pi}\,W(\omega)\,
S_{yy}^{\mathrm{res},+}(\omega)/|\chi(\omega)|^2
}{
4\int_{\mathcal B}\frac{\dd\omega}{2\pi}\,W(\omega)
},
\label{eq:Dtau_band}
\end{equation}
with $W(\omega)\ge 0$ chosen, for instance, to emphasize the band where the residual spectrum is
best controlled.
Equations~(\ref{eq:Dtau_point}) and (\ref{eq:Dtau_band}) are intentionally model-independent at
the spectrum level. They make no assumptions about the microscopic origin of the stochastic term
beyond stationarity. By contrast, the deterministic Fisher/Cram\'er--Rao bounds in
Sec.~\ref{sec:deterministic} additionally assume approximately Gaussian processed noise in the
analysis band; when strong non-Gaussian tails or obvious nonstationarity remain, the safer
quantity to quote is the spectrum-level bound in Eq.~(\ref{eq:pointwise_bound}) rather than the
ideal matched-filter error bar.

At microscopic level, however, many candidate models are first written in terms of a
force-density or momentum-diffusion kernel rather than a torque directly. For a torsional
mode about $\hat z$ one has schematically
\begin{align}
\tau_z(t)&=\int \dd^3r\, [\mathbf r\times \mathbf f(\mathbf r,t)]_z,
\nonumber\\
S_{\tau\tau}^{\scg}(\omega)&=\int \dd^3r\,\dd^3r'\,
\epsilon_{zij}r_i\,\epsilon_{zkl}r'_k\,
S_{f_j f_l}^{\scg}(\mathbf r,\mathbf r';\omega).
\label{eq:torque_formfactor}
\end{align}
with repeated Cartesian indices summed, or the corresponding discrete-mass sum. A
force-diffusion model therefore maps to $D_\tau$ only after the geometry/form-factor
integral implied by Eq.~(\ref{eq:torque_formfactor}) is carried out; in a rigid-body
lever-arm estimate this reduces schematically to
$D_\tau\sim \mathcal G_{\mathrm{geom}}\,l_{\mathrm{eff}}^2D_F$, where
$\mathcal G_{\mathrm{geom}}=\mathcal O(1)$ and $D_F$ is the white diffusion coefficient of the
generalized force projected onto the relevant translational mode, defined by
$S_{FF}(\omega)=4D_F$ and therefore carrying units $\mathrm{N^2\,s}$.

This intermediary role appears in Oppenheim-type classical-quantum gravity and in its
weak-field elaborations. There the fundamental diffusion is formulated for classical gravitational
variables, and the induced torque spectrum of a specific torsion mode depends on geometry,
coarse-graining, and readout conventions rather than on a universal setup-independent number
\cite{Oppenheim2023,Layton2023}. There is therefore no single postquantum ``target'' value of
$D_\tau$ that can be quoted independently of the apparatus; producing one requires exactly the
projection in Eq.~(\ref{eq:torque_formfactor}) for the suspension and source geometry at hand.
Our $D_\tau$ is the experiment-facing quantity that any concrete theory must map onto before
comparison with data, and a full Oppenheim-to-torsion mapping for realistic pendula is left to
future work. The same theory-to-experiment workflow already appears in the collapse-model
literature, where rotational-noise data have been translated into geometry-dependent CSL
exclusions for both tabletop and space-based platforms \cite{Altamura2024,Altamura2025}.

\subsection{Equivalent torque spectra as the natural comparison space}
For the stochastic problem we divide the measured angle spectrum by
$|\chi|^2$ and work in equivalent torque units,
\begin{equation}
S_{\tau\tau}^{\eqv}(\omega)\equiv \frac{S_{yy}(\omega)}{|\chi(\omega)|^2}.
\label{eq:Seqv}
\end{equation}
Equation~(\ref{eq:Seqv}) collapses any torque-like source into a directly comparable quantity.
In practice, one should divide by the magnitude of the calibrated response actually used in the
subtraction; for a structural-loss calibration this means $|\chi|\rightarrow|\chi_\phi|$.
White viscous thermal noise becomes a flat line [Eq.~(\ref{eq:viscous})], structural damping
becomes the $1/\omega$ spectrum in Eq.~(\ref{eq:structural_torque}), and a white semiclassical
diffusion term is then a constant $4D_\tau$. Accordingly, Fig.~\ref{fig:torque_asd} below is
arguably more informative for semiclassical-gravity constraints than the raw angle ASD.

\section{Quantitative benchmarks}
\label{sec:benchmarks}

\subsection{Representative room-temperature Cavendish benchmark}
The numbers in this subsection are an illustrative scale-setting model, not a universal prediction
for all torsion balances. They are chosen to make the conversion between mechanical parameters,
angle spectra, equivalent torque spectra, SQL benchmarks, and white torque-diffusion levels
transparent.

We begin with a benchmark chosen to display the scalings explicitly:
\begin{align}
I&=5.0\times10^{-5}\,\mathrm{kg\,m^2}, &
\kappa&=1.0\times10^{-7}\,\mathrm{N\,m/rad}, \nonumber\\
Q&=5.0\times10^3, &
T&=300\,\mathrm{K}.
\label{eq:benchmark_params}
\end{align}
The corresponding resonance is
\begin{align}
f_0&=\frac{\omega_0}{2\pi}
=\frac{1}{2\pi}\sqrt{\frac{\kappa}{I}}
\simeq 7.12\,\mathrm{mHz},
\nonumber\\
\theta_{\zpf}&\simeq 4.86\times10^{-15}\,\mathrm{rad}.
\label{eq:benchmark_derived}
\end{align}
For the simple two-end-mass picture one may regard this as $I=2ml^2$ with
$m=10\,\mathrm{g}$ and $l=5\,\mathrm{cm}$, but in the rest of the paper $I$ is always treated
as the measured mechanical parameter rather than inferred from a point-mass model.

The resonant susceptibility is
\begin{equation}
|\chi(\omega_0)|
=
\frac{Q}{I\omega_0^2}
=
5.0\times10^{10}\,\mathrm{rad/(N\,m)}.
\label{eq:chi_res}
\end{equation}
To define a reference quantum-limited detector, we choose white imprecision and white backaction
spectra saturating Eq.~(\ref{eq:quantum_ineq}) and touching the SQL at resonance:
\begin{equation}
S_{\theta\theta}^{\imp}=\hbar|\chi(\omega_0)|,
\qquad
S_{\tau\tau}^{\ba}=\frac{\hbar}{|\chi(\omega_0)|}.
\label{eq:ref_quantum}
\end{equation}
Numerically,
\begin{align}
\sqrt{S_{\theta\theta}^{\imp}}
&\simeq
2.30\times10^{-12}\,\mathrm{rad}/\sqrt{\mathrm{Hz}},
\nonumber\\
\sqrt{S_{\tau\tau}^{\ba}}
&\simeq
4.59\times10^{-23}\,\mathrm{N\,m}/\sqrt{\mathrm{Hz}}.
\label{eq:ref_quantum_num}
\end{align}
The measurement-added SQL at resonance is therefore
\begin{equation}
\sqrt{S_{\theta\theta}^{\SQL}(\omega_0)}
=
\sqrt{2\hbar|\chi(\omega_0)|}
\simeq
3.25\times10^{-12}\,\mathrm{rad}/\sqrt{\mathrm{Hz}}.
\label{eq:sql_num}
\end{equation}
Equivalently,
$S_{\theta\theta}^{\SQL}(\omega_0)=4\theta_{\zpf}^2Q/\omega_0$,
which leaves $\theta_{\zpf}$ as a single-oscillator benchmark even though the
main text works directly with calibrated spectra.

By contrast, the one-sided viscous thermal torque PSD is
\begin{equation}
S_{\tau\tau}^{\thm,\mathrm{visc}}
=
4\kb T I\gamma
=
7.41\times10^{-30}\,\mathrm{N^2\,m^2/Hz},
\label{eq:thermal_torque_num}
\end{equation}
corresponding to a resonant angle ASD
\begin{equation}
\sqrt{S_{\theta\theta}^{\thm,\mathrm{visc}}(\omega_0)}
=
1.36\times10^{-4}\,\mathrm{rad}/\sqrt{\mathrm{Hz}}.
\label{eq:thermal_angle_num}
\end{equation}
Thus the benchmark thermal floor exceeds the measurement-added SQL by a factor
$1.36\times10^{-4}/(3.25\times10^{-12})\simeq4.2\times10^7$ in ASD, i.e. about $7.6$ orders of
magnitude (or $\sim1.8\times10^{15}$ in PSD). This large gap is precisely why current
torsion-balance experiments are not usually limited by quantum backaction, even though the SQL
remains the correct asymptotic benchmark once classical torque noise is reduced.

\begin{figure}[t]
\centering
\includegraphics[width=0.98\linewidth]{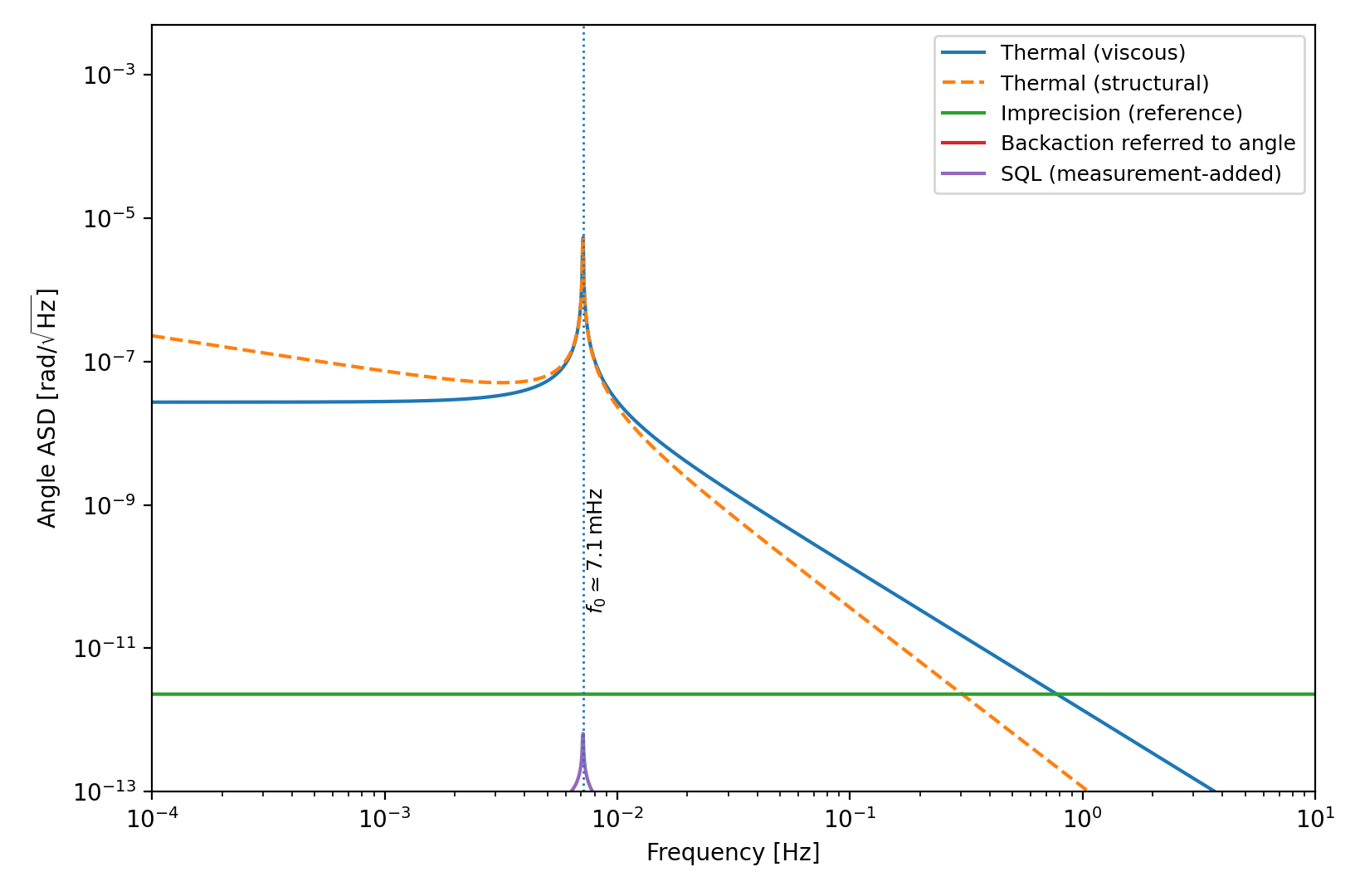}
\caption{Benchmark angle ASDs for the representative Cavendish parameters in
Eq.~(\ref{eq:benchmark_params}). The thermal curves are shown for both viscous and structural
damping. The imprecision and backaction curves are the reference pair of
Eq.~(\ref{eq:ref_quantum}), tuned to touch the measurement-added SQL at resonance. The values at
$f_0\simeq7.1\,\mathrm{mHz}$ are numerically consistent with
Eqs.~(\ref{eq:ref_quantum_num})--(\ref{eq:thermal_angle_num}).}
\label{fig:angle_asd}
\end{figure}

Figure~\ref{fig:angle_asd} shows the linear calibrated angle-ASD budget, including both viscous and
structural thermal models. The corresponding viscous thermal diffusion scale is
\begin{equation}
D_\tau^{\mathrm{bench}}=\frac{S_{\tau\tau}^{\thm,\mathrm{visc}}}{4}
=
1.85\times10^{-30}\,\mathrm{N^2\,m^2\,s}.
\label{eq:Dtau_bench}
\end{equation}
Equation~(\ref{eq:Dtau_bench}) is not an experimental semiclassical-gravity bound; it is the
illustrative stochastic torque level that a residual analysis would need to beat on this
representative room-temperature platform.

An experimentalist should read Fig.~\ref{fig:angle_asd} as an angle-domain diagnostic. A
narrowband deterministic torque template at frequency $\omega_m$ is constrained by taking the
local ordinate $\sqrt{S_{yy}(\omega_m)}$ from the relevant processed spectrum and inserting it in
Eq.~(\ref{eq:narrowband}). The figure also shows why reducing the optical imprecision alone does
not automatically improve a millihertz gravitational test: if the observed local floor is thermal
or environmental, the matched-filter gain is controlled by that classical floor until it is
subtracted or physically reduced.

\subsection{A public stochastic benchmark from the 2025 torsion-balance search}
A more directly experimental benchmark is available from the recent torsion-balance search of
Yan \emph{et al.}, who reported an ultralow-frequency optomechanical platform with
$f_0=0.6\,\mathrm{mHz}$, quality factor exceeding $5\times10^4$, and an angle sensitivity of
$0.3\,\mu\mathrm{rad}/\sqrt{\mathrm{Hz}}$ at $2.5\,\mathrm{mHz}$ \cite{Yan2025}. Using the
reported torsional moment of inertia $I=0.14\,\mathrm{kg\,m^2}$ for the same rotational mode
\cite{Yan2025,Prokhorov2024}, the susceptibility magnitude at $2.5\,\mathrm{mHz}$ is
\begin{equation}
|\chi(2.5\,\mathrm{mHz})|
\simeq
3.07\times10^4\,\mathrm{rad/(N\,m)}.
\label{eq:chi_yan}
\end{equation}
Treating the published $0.3\,\mu\mathrm{rad}/\sqrt{\mathrm{Hz}}$ as a conservative bound on the
\emph{total} observed floor, the corresponding equivalent torque ASD is
\begin{equation}
\sqrt{S_{\tau\tau}^{\eqv}}
=
\frac{0.3\times10^{-6}\,\mathrm{rad}/\sqrt{\mathrm{Hz}}}
{|\chi(2.5\,\mathrm{mHz})|}
\simeq
9.77\times10^{-12}\,\mathrm{N\,m}/\sqrt{\mathrm{Hz}},
\label{eq:yan_torque_asd}
\end{equation}
which implies the conservative observed-floor white-diffusion bound
\begin{equation}
D_\tau^{\mathrm{Yan}}
\lesssim
\frac{S_{yy}^{\mathrm{obs}}}{4|\chi|^2}
\simeq
2.38\times10^{-23}\,\mathrm{N^2\,m^2\,s}.
\label{eq:yan_Dtau}
\end{equation}
Equation~(\ref{eq:yan_Dtau}) is conservative because it uses a quoted observed
sensitivity floor rather than a fully subtracted residual semiclassical-noise spectrum. We use this single public sensitivity point
as a reference worked example; a full reanalysis of the published spectrum would
employ Eq.~(\ref{eq:Dtau_band}) over the full residual band. Using only the minimum
published quality factor, the quoted resonance frequency, and a room-temperature estimate gives
a reference
viscous thermal scale for the same platform,
\begin{equation}
D_{\tau,\thm}^{\mathrm{Yan,min}}
\equiv
\kb T\,I\frac{\omega_0}{Q}
\simeq
4.4\times10^{-29}\,\mathrm{N^2\,m^2\,s},
\label{eq:yan_thermal_Dtau}
\end{equation}
where we used $Q=5\times10^4$, $f_0=0.6\,\mathrm{mHz}$, and $T=300\,\mathrm{K}$.
This is about $5\times10^5$ times smaller than Eq.~(\ref{eq:yan_Dtau}), consistent with Yan
\emph{et al.}'s statement that thermal noise was not the dominant observed floor near
$2.5\,\mathrm{mHz}$ \cite{Yan2025}. The public conservative bound therefore does not reduce to the
platform's thermal floor rewritten in different units.

As a theory-facing orientation point, Eq.~(\ref{eq:yan_Dtau}) may also be read as a bound on a
white generalized-force diffusion,
\begin{equation}
D_F \lesssim \frac{D_\tau^{\mathrm{Yan}}}{l_{\mathrm{eff}}^2}
\simeq
2.4\times10^{-19}
\left(\frac{l_{\mathrm{eff}}}{1\,\mathrm{cm}}\right)^{-2}
\mathrm{N^2\,s},
\label{eq:yan_force_diffusion}
\end{equation}
for an effective lever arm $l_{\mathrm{eff}}$. Device-specific theory comparisons must then
supply the appropriate form factor of Eq.~(\ref{eq:torque_formfactor}). Rotational-noise
analyses have already been mapped into specific collapse-model parameter bounds. In particular,
Altamura \emph{et al.} showed that a tabletop rotational-noise analysis can constrain CSL near
$\lambda\sim10^{-9}\,\mathrm{s^{-1}}$ at $r_C=10^{-4}\,\mathrm{m}$ and that geometry
optimization can substantially tighten the bound, while rotational noise from LISA Pathfinder
strengthens the constraints further \cite{Altamura2024,Altamura2025}. We do not translate those
results into a universal $D_\tau$, because the mapping depends on the detailed mass-density form
factor of the experimental geometry, but they illustrate exactly why the intermediary
experimental quantity should be a calibrated torque spectrum rather than a theory-specific
parameter. The same logic underlies direct semiclassical-source proposals such as optomechanical
tests of the field generated by a quantum superposition and Schr\"odinger--Newton-induced
dephasing analyses: the theory is first posed in native source parameters, while the experiment
responds to a geometry-dependent force or torque observable that must be projected onto the
measured mode \cite{Carlesso2019,Grossardt2021}.

\begin{figure*}[t]
\centering
\includegraphics[width=0.76\linewidth]{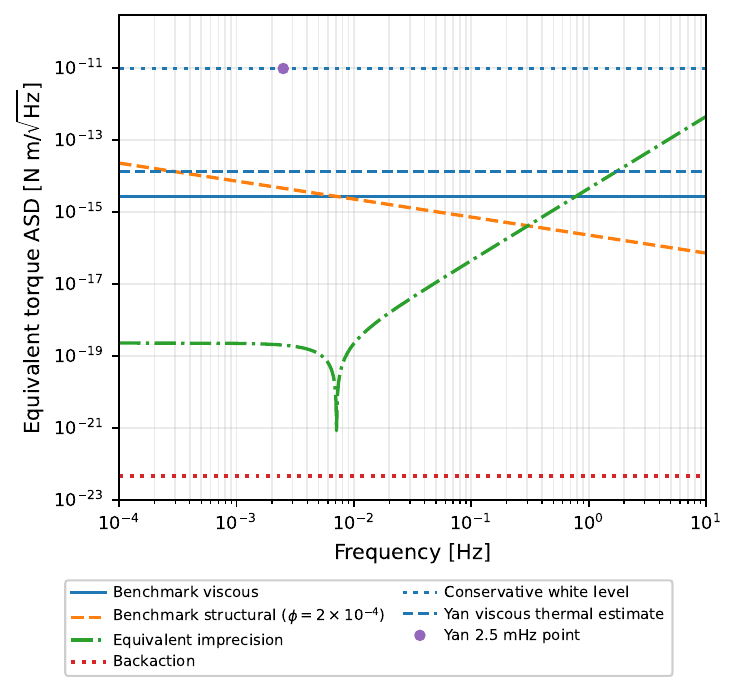}
\caption{Equivalent torque ASDs on common axes. The benchmark curves show, in equivalent
torque units, the representative room-temperature viscous thermal, structural thermal,
equivalent-imprecision, and backaction levels for Eq.~(\ref{eq:benchmark_params}); the
structural-loss curve uses $\phi=2\times10^{-4}$ only as an illustration of the red
$1/\sqrt{f}$ scaling in equivalent torque units. The point marks the public sensitivity quoted
by Yan \emph{et al.} at $2.5\,\mathrm{mHz}$, converted by
Eq.~(\ref{eq:yan_torque_asd}). The dashed horizontal line shows the illustrative viscous
thermal level inferred from the published $f_0$, $Q$, and $I$ via
Eq.~(\ref{eq:yan_thermal_Dtau}); the dotted horizontal line shows the conservative white level
obtained by interpreting the public point as a frequency-independent floor. The common-axis
overlay places benchmark and platform-specific quantities in the same equivalent-torque units; it
does not model a single common suspension.}
\label{fig:torque_asd}
\end{figure*}

Figure~\ref{fig:torque_asd} places the representative room-temperature benchmark
curves and the Yan-specific reference on common axes. The benchmark curves set the scale of
measurement-added and thermal torque noise for Eq.~(\ref{eq:benchmark_params}), while the
public Yan point and the illustrative viscous thermal reference of
Eq.~(\ref{eq:yan_thermal_Dtau}) are overlaid in the same equivalent-torque units. The overlay
is intended as a unit-consistent comparison between benchmark and platform-specific quantities,
not as a statement that both datasets arise from a single suspension model. At sufficiently low
frequency, the equivalent torque ASD of a structurally damped fiber rises as $1/\sqrt{f}$ and
can be misread as an exotic stochastic drive if the suspension model is not independently
characterized.

For experimental design, Fig.~\ref{fig:torque_asd} is the more direct plot for stochastic models.
A candidate model predicting a torque spectrum $S_{\tau\tau}^{\mathrm{model}}(\omega)$ should be
plotted against the residual equivalent-torque spectrum rather than against the raw angle ASD. A
flat model can be summarized by $D_\tau$, but a colored model is constrained only where its full
frequency dependence lies below the residual upper band. The figure also shows that an observed
public floor and a thermal estimate for the same platform answer different questions: the former
is an experimental observed-floor bound, while the latter is a suspension-noise scale inferred
from a simplified damping model.

\subsection{Classical-source benchmarks from E\"ot-Wash}
The strongest existing torsion-balance tests of the WEP constrain composition-dependent
differential accelerations at the part-in-$10^{13}$ level \cite{Wagner2012WEP}, while
inverse-square-law searches exclude gravitational-strength Yukawa interactions down to
tens of micrometers \cite{Kapner2007,Tan2020}. We do not attempt to re-extract those limits from
raw spectra here. Instead, the point relevant for semiclassical gravity is simpler: any candidate
classical-source modification must reproduce the same null results when projected onto the
corresponding torque templates $\Lambda_x(t)$. In other words, the deterministic matched-filter
framework already built into standard torsion-balance null tests is exactly the correct language
for the deterministic sector of semiclassical-gravity phenomenology.

\section{Page--Geilker and MEP as matched-filter applications}
\label{sec:pagekent}
This section is an interpretive application of the general template formalism, not a claim that
Page--Geilker branch discrimination and Fedida--Kent-style MEP tests are physically equivalent.
They differ in state-preparation logic and in the physical sourcing rule being tested. The common
structure appears only after each competing prediction has been projected onto the torque that the
torsion balance would see. For any two candidate sourcing rules $A$ and $B$, let
\begin{equation}
\Delta\tau(t)\equiv \tau_A(t)-\tau_B(t).
\label{eq:delta_tau_rules}
\end{equation}
The optimal Gaussian discrimination statistic is governed by
\begin{equation}
\mathrm{SNR}_{A/B}^2
=
4\int_0^\infty\frac{\dd\omega}{2\pi}
\frac{|\chi(\omega)\Delta\tau(\omega)|^2}{S_{yy}(\omega)}.
\label{eq:SNR_rule_general}
\end{equation}
The narrowband formula below is the single-line version of Eq.~(\ref{eq:SNR_rule_general}).
Thus Page--Geilker and MEP tests reduce to the same criterion only after their different physical
sourcing assumptions have been converted into the torque difference $\Delta\tau$.

\subsection{What Page--Geilker excludes}
Consider two macroscopically distinct source configurations $|L\rangle$ and $|R\rangle$ that
produce torques $\pm\tau_0(t)$ on the balance. A naive M{\o}ller--Rosenfeld mean-field theory
sources gravity from the expectation value of the mass distribution. For the symmetric
superposition $(|L\rangle+|R\rangle)/\sqrt{2}$ this implies an approximately vanishing
single-run torque, whereas branch-correlated behavior would produce responses $\pm\tau_0$.
The Page--Geilker result is therefore often interpreted as excluding the simplest
deterministic, no-collapse mean-field sourcing rule on individual runs \cite{PageGeilker1981,Giulini2022}.
It does \emph{not} by itself exclude every semiclassical model: collapse-assisted or stochastic
versions can evade the contradiction by producing definite branchwise source configurations or
additional diffusion \cite{Giulini2022,Oppenheim2023}.

\subsection{A quantitative branch-discrimination criterion}
If one compares a branch-correlated torque $\pm\tau_0\cos\omega_m t$ to an averaged or vanishing
mean-field response at the same modulation frequency, the relevant signal-to-noise ratio is just
the narrowband matched-filter expression
\begin{equation}
\mathrm{SNR}_{\mathrm{branch}}
\simeq
\frac{|\chi(\omega_m)|\,\tau_0}{\sqrt{S_{yy}(\omega_m)B_w}}.
\label{eq:SNR_branch}
\end{equation}
If the sign or phase of the branch-correlated response is not fixed a priori and one performs a
two-sided test over both quadratures, the corresponding detection threshold acquires the same
$\sqrt{2}$ factor discussed below Eq.~(\ref{eq:narrowband}). For the representative thermal
benchmark of Sec.~\ref{sec:benchmarks}, choosing
$\omega_m=\omega_0$, a rectangular window, and $T_{\mathrm{int}}=1$ day yields
\begin{equation}
\tau_{0,\mathrm{min}}
\sim
\frac{\sqrt{S_{yy}(\omega_0)/T_{\mathrm{int}}}}{|\chi(\omega_0)|}
\approx
9\times10^{-18}\,\mathrm{N\,m}
\label{eq:taumin_branch}
\end{equation}
for unit signal-to-noise if the observed floor is thermal. The numerical value follows by using
$|\chi(\omega_0)|=Q/\kappa=5.0\times10^{10}\,\mathrm{rad/(N\,m)}$ and
$\sqrt{S_{yy}(\omega_0)}=1.36\times10^{-4}\,\mathrm{rad}/\sqrt{\mathrm{Hz}}$ from
Eq.~(\ref{eq:thermal_angle_num}), so that the coherent one-day angle uncertainty is
$1.36\times10^{-4}/\sqrt{86400}\simeq4.6\times10^{-7}\,\mathrm{rad}$. Equation~(\ref{eq:taumin_branch})
should not be overinterpreted: it is a benchmark calculation showing how a projected branch
template would be tested, not a proposal for an immediately feasible macroscopic superposition
experiment or a geometry-specific Page--Geilker/MEP design.

\subsection{Proper and improper mixtures}
Fedida and Kent recently emphasized that M{\o}ller--Rosenfeld semiclassical gravity violates a
weak form of the mixture equivalence principle: proper and improper mixtures with the same
density matrix need not be gravitationally equivalent in such a theory \cite{FedidaKent2025}.
The torsion-balance formulation converts that statement into a comparison of torque templates.

Define
\begin{align}
\rho_{\mathrm{proper}}
&=
\frac{1}{2}\bigl(|L\rangle\langle L|+|R\rangle\langle R|\bigr), \\
|\Psi\rangle_{SA}
&=
\frac{1}{\sqrt{2}}\bigl(|L\rangle_S|0\rangle_A+|R\rangle_S|1\rangle_A\bigr), \\
\rho_{\mathrm{improper}}
&=
\mathrm{Tr}_A\!\left(|\Psi\rangle\langle\Psi|\right),
\end{align}
so that $\rho_{\mathrm{proper}}=\rho_{\mathrm{improper}}$ as density operators on the source
subsystem.
If gravity respects mixture equivalence, then an unconditional torsion-balance measurement cannot
distinguish the two preparations. Operationally, the distinction is between two classes of
sourcing rules. A reduced-state rule assigns an unconditional torque functional
$\tau_{\mathrm{red}}(t)=\mathcal T[\rho_S](t)$ and therefore necessarily obeys
$\tau_{\mathrm{red}}[\rho_{\mathrm{proper}}]=\tau_{\mathrm{red}}[\rho_{\mathrm{improper}}]$.
A branchwise rule instead assigns $\tau_b(t)=b\,\tau_0(t)$ with $b=\pm1$ on each run and only
vanishes after ensemble averaging. Any observable difference between the two preparations
therefore presupposes a nonstandard gravitational sourcing rule that is \emph{not} a function
solely of the reduced density operator $\rho_S$; ordinary linear quantum mechanics coupled only
through $\rho_S$ would predict no unconditional distinction. In one representative
MEP-violating class of mean-field models discussed by Fedida and Kent, the improper
preparation is taken to source the averaged mass distribution and hence yields a near-zero
unconditional torque, whereas the proper mixture yields a run-by-run bimodal response
$\pm\tau_0$. The discrimination criterion is then again Eq.~(\ref{eq:SNR_branch}); the
only new element is the state-preparation requirement.

Two remarks follow. First, the original Page--Geilker protocol is related to but not
identical with a full MEP test, because it does not compare proper and improper mixtures with the
same reduced density matrix. Second, the bottleneck in a future MEP experiment is not the
statistical analysis, which follows the same matched-filter criterion, but the source preparation: one must realize a
source whose gravitational torque is large enough to satisfy Eq.~(\ref{eq:SNR_branch}) while
maintaining control over uncontrolled decoherence and classical randomness. The torsion-balance
formalism developed here isolates that requirement.

\section{Experimental design implications}
\label{sec:discussion}
The unified deterministic--stochastic framework yields four experimental design implications.

First, the public data product directly used by the bounds in Secs.~\ref{sec:deterministic} and \ref{sec:diffusion} is a
calibrated residual \emph{equivalent torque} spectrum rather than only a raw angle ASD. In the
absence of strong imprecision--backaction correlations, the residual can be
written schematically as
\begin{align}
S_{\tau\tau}^{\eqv,\mathrm{res}}(\omega)
\simeq {}&
\frac{S_{yy}^{\mathrm{obs}}(\omega)}{|\chi(\omega)|^2}
-S_{\tau\tau}^{\thm}(\omega)
-\sum_j |K_j(\omega)|^2 S_{x_jx_j}(\omega)
\nonumber\\
&-\frac{S_{\theta\theta}^{\imp}(\omega)}{|\chi(\omega)|^2}
-S_{\tau\tau}^{\ba}(\omega).
\label{eq:residual_recipe}
\end{align}
where $K_j\equiv \partial\tau/\partial x_j$ converts each auxiliary environmental channel into
an equivalent torque contribution. Equation~(\ref{eq:residual_recipe}) is the diagonal version.
If the auxiliary channels are correlated---as they often are when several monitors observe the
same Newtonian-noise field---the subtraction should instead use the full cross-spectral matrix,
\begin{equation}
S_{\tau\tau}^{\mathrm{aux}}(\omega)
=
\sum_{i,j} K_i(\omega)K_j^\ast(\omega)S_{x_i x_j}(\omega).
\label{eq:aux_cross_matrix}
\end{equation}
Thus the separation of noises is a calibrated multichannel inference problem, not an algebraic
identity. Low-frequency excess backgrounds --- seismic tilt, thermal drift, gravity-gradient
fluctuations driven, for example, by atmospheric pressure fields or slow local mass redistribution
(including hydrological changes), magnetic pickup, and other $1/f$-like technical noise --- enter
through these auxiliary channels or, if left unsubtracted, remain part of the residual itself
\cite{Harms2015,MohantaShikano2026ANN}. When the candidate stochastic signal is itself colored
toward low frequency, independent barometric, gravimetric, tilt, or hydrological monitors provide
independent constraints, because atmospheric and local-mass Newtonian-noise channels can otherwise
be spectrally degenerate with the target signal. If this degeneracy cannot be broken, the correct
claim is a conservative bound based on the observed total floor rather than a separated residual.
Ref.~\cite{MohantaShikano2026ANN} gives a complementary analysis of atmospheric Newtonian-noise
propagation for torsion-balance $G$ measurements. Such a spectrum can be inserted immediately into
Eqs.~(\ref{eq:Dtau_point}) and (\ref{eq:Dtau_band}).

The single formula covers all auxiliary channels. For example, $x_j$ may be platform tilt,
barometric pressure, a gravity-gradient monitor, a hydrological proxy, magnetic field, charge or
electric-field monitor, temperature, or a deliberately injected calibration line. What matters is
not the label of the channel but the measured cross-spectral matrix and the torque calibration
$K_j(\omega)$ used to map that channel into equivalent torque units.

Second, any published residual should be accompanied by the metadata needed to reproduce it:
independent identification of $I$, $\kappa$, and $Q$ from ringdown and driven transfer-function
measurements; the window, ENBW, and leakage treatment entering the spectrum estimate; and the
auxiliary calibrations $K_j$ and cross spectra used to subtract the dominant environmental
channels. Appendix~\ref{app:publication} summarizes this recipe compactly.

Third, structural damping must be characterized independently. If one extrapolates a
room-temperature torsion balance with the wrong loss model into the millihertz band, the
resulting equivalent torque floor can be wrong by orders of magnitude. A minimal characterization
program is to measure the quality factor as a function of amplitude and temperature, then compare
the inferred $S_{\tau\tau}^{\eqv}$ with the white prediction of Eq.~(\ref{eq:viscous}) and the
$1/\omega$ prediction of Eq.~(\ref{eq:structural_torque}).

Fourth, quantum-limited readout remains relevant even if present experiments are
dominated by classical torque noise. The reason is not that the SQL itself is the target of the
current semiclassical-gravity bounds, but that any future attempt to sharpen deterministic branch
discrimination or stochastic diffusion constraints will eventually require a readout whose
imprecision, backaction, and classical environmental couplings have been separately identified.
Appendix~\ref{app:variational} records a generic two-channel variational or speed-meter-like
strategy. In the present paper it serves as an extension of the classical-noise model, not as a
replacement for it.

\section{Conclusions}
This paper gives a weak-field operational map from calibrated torsion-balance spectra to
constraints on deterministic torque templates and additional stationary stochastic torque noise.
The result is not a direct test of the full relativistic semiclassical Einstein equation. It is a
Newtonian-limit measurement interface: a candidate sourcing rule is projected to a torque
waveform or torque-noise spectrum, and that object is compared with the calibrated angular data.

The formal framework consists of two reusable equations. For deterministic effects, any torque
template $\tau_x(t)=g_x\Lambda_x(t)$ is bounded by the matched-filter Fisher information in
Eq.~(\ref{eq:fisher_general}), with the finite-time, window-aware form in
Eq.~(\ref{eq:narrowband}). For stochastic effects, any additional stationary torque noise is
bounded by the residual equivalent-torque spectrum in Eq.~(\ref{eq:pointwise_bound}). The scalar
coefficient $D_\tau$ is only the Markovian white-noise compression of this spectrum; colored or
non-Markovian models must be compared with the full frequency-dependent residual upper band.
Page--Geilker branch tests and Fedida--Kent mixture-equivalence tests are interpretive
applications of this framework once their physically different sourcing rules are expressed as
torque templates, not a claim of physical equivalence between the two experiments.

The numerical results have three distinct roles. The representative room-temperature Cavendish
calculation is an illustrative benchmark: it shows that, for the chosen parameters, the resonant
thermal angle ASD exceeds the measurement-added SQL by about $4.2\times10^7$ in amplitude and
that the corresponding viscous thermal white-diffusion scale is
$D_\tau^{\mathrm{bench}}\simeq1.9\times10^{-30}\,\mathrm{N^2\,m^2\,s}$. This benchmark is useful for
scale-setting but is not an experimental semiclassical-gravity limit. The Yan \emph{et al.}
conversion is different: the published $0.3\,\mu\mathrm{rad}/\sqrt{\mathrm{Hz}}$ point at
$2.5\,\mathrm{mHz}$ gives a conservative observed-floor white-diffusion bound of order
$10^{-23}\,\mathrm{N^2\,m^2\,s}$, not a fully subtracted residual semiclassical-noise constraint.
The inferred Yan thermal scale near $10^{-29}\,\mathrm{N^2\,m^2\,s}$ is a suspension-noise
orientation point based on the published $f_0$, $Q$, and $I$, and should not be conflated with the
observed-floor bound.

Figures~\ref{fig:angle_asd} and \ref{fig:torque_asd} show how these numbers should be used by
experimentalists. The angle-ASD plot identifies the local floor that enters the finite-time
matched-filter uncertainty for a deterministic template. The equivalent-torque plot is the more
natural comparison space for stochastic gravitational models: a white model can be summarized by a
horizontal $4D_\tau$ level, whereas a colored model must be compared frequency by frequency with
the residual equivalent-torque spectrum. In both cases the design target is an operational band,
chosen where the calibrated Fisher kernel is large and where Newtonian, thermal, technical, and
quantum readout noises can be separated.

Application of these expressions requires calibrated residual spectra, suspension-loss
characterization, and auxiliary-channel transfer functions, including cross-spectral information
when several monitors observe the same Newtonian-noise field. With those ingredients,
torsion-balance data can be reported in a form that supports direct reuse in deterministic
template tests and stochastic semiclassical-gravity searches. Deterministic Newtonian-noise
sensing, source classification, recovery-time analysis, and inverse reconstruction are natural
extensions of the same formalism, but they require geometry-specific templates and auxiliary
channels and are best developed as a separate follow-up study.

\begin{acknowledgments}
The authors thank Adrian Kent for inspiring discussions on semiclassical gravity.
This work was partially supported by JST ASPIRE (No.~JPMJAP2339) and the JST SPRING Program (No.~JPMJSP2124).
\end{acknowledgments}

\appendix

\section{One-sided spectral conventions and the structural-loss FDT}
\label{app:fdt}
Because the manuscript combines quantum measurement noise, thermal noise, and stochastic
semiclassical drive noise, the spectral convention must be fixed once and for all.
For a real stationary process $x(t)$ we define the one-sided PSD $S_{xx}(\omega)$ for $\omega>0$
by
\begin{equation}
\langle x(\omega)x^\ast(\omega')\rangle
=
2\pi\,\delta(\omega-\omega')\,S_{xx}(\omega),
\qquad
\omega,\omega'>0,
\end{equation}
and the ASD is $\sqrt{S_{xx}}$.

With the Fourier convention used in the main text and the structural-loss response
$\chi_\phi(\omega)=[\kappa(1-\ii\phi)-I\omega^2]^{-1}$, the one-sided fluctuation--dissipation
relation for the angle coordinate is
\begin{equation}
S_{\theta\theta}^{\thm,\mathrm{str}}(\omega)=\frac{4\kb T}{\omega}\operatorname{Im}\chi_\phi(\omega).
\end{equation}
Since
\begin{equation}
\operatorname{Im}\chi_\phi(\omega)
=
\frac{\kappa\phi}{(\kappa-I\omega^2)^2+(\kappa\phi)^2},
\end{equation}
one immediately obtains Eq.~(\ref{eq:structural_torque}) after dividing by
$|\chi_\phi|^2$.
The corresponding double-sided symmetrized expressions differ by the familiar factor of two.

\section{White torque diffusion and angular-momentum variance}
\label{app:dtau_diffusion}
This appendix fixes the stochastic-process meaning of $D_\tau$. Let the stochastic torque in the
white Markovian limit be denoted by $\xi_\tau(t)$ and write the angular-momentum equation as
\begin{equation}
\dot L(t)=\xi_\tau(t),
\qquad
\langle \xi_\tau(t)\xi_\tau(t')\rangle=2D_\tau\delta(t-t'),
\label{eq:white_torque_corr}
\end{equation}
where the correlation function is understood as the double-sided time-domain convention. Then
\begin{align}
\left\langle [L(t)-L(0)]^2\right\rangle
&=
\int_0^t\dd t_1\int_0^t\dd t_2\,
\langle \xi_\tau(t_1)\xi_\tau(t_2)\rangle
\nonumber\\
&=2D_\tau t.
\label{eq:L_diffusion}
\end{align}
Thus $D_\tau$ is the coefficient controlling random-walk growth of angular-momentum variance.
The double-sided white PSD corresponding to Eq.~(\ref{eq:white_torque_corr}) is
$2D_\tau$, and the one-sided PSD used in the main text is therefore
\begin{equation}
S_{\tau\tau}^{\mathrm{1s}}(\omega)=4D_\tau,
\qquad \omega>0.
\label{eq:Dtau_one_sided_app}
\end{equation}
If the torque noise is colored, the angular-momentum variance is instead determined by the full
correlation function or full spectrum; no single $D_\tau$ captures the process.

\section{Finite-time matched filtering}
\label{app:finite_time}
For completeness we sketch the finite-time normalization leading to
Eq.~(\ref{eq:narrowband}) in the single-tone case. Consider data
\begin{equation}
y(t)=A\cos(\omega_m t+\varphi)+n(t),
\qquad
0<t<T_{\mathrm{int}},
\end{equation}
where $n(t)$ is stationary Gaussian noise with one-sided PSD
$S_{yy}(\omega_m)$ that is approximately constant across the bandwidth of the analysis window.
For a chosen window $w(t)$, define the normalized quadrature estimator
\begin{equation}
\hat A
\equiv
\frac{2}{\int_0^{T_{\mathrm{int}}}\dd t\, w(t)}
\int_0^{T_{\mathrm{int}}}\dd t\, w(t)\,y(t)\cos(\omega_m t+\varphi).
\end{equation}
Averaging $\cos^2$ over many cycles gives $\langle \hat A\rangle \simeq A$. For the same
many-cycle approximation, the variance of the estimator is
\begin{align}
\mathrm{Var}(\hat A)
\simeq{}&
\frac{4}{\left[\int_0^{T_{\mathrm{int}}}\dd t\, w(t)\right]^2}
\left(\frac{S_{yy}(\omega_m)}{2}\right)
\nonumber\\
&\times
\int_0^{T_{\mathrm{int}}}\dd t\, w^2(t)\cos^2(\omega_m t+\varphi),
\end{align}
and hence
\begin{equation}
\mathrm{Var}(\hat A)
\simeq
S_{yy}(\omega_m)
\frac{\int_0^{T_{\mathrm{int}}}\dd t\, w^2(t)}
{\left[\int_0^{T_{\mathrm{int}}}\dd t\, w(t)\right]^2}
=
S_{yy}(\omega_m)B_w.
\end{equation}
With the one-sided convention used in the main text, the factor $1/2$ from the many-cycle
average $\langle\cos^2\rangle=1/2$ is exactly cancelled by the factor relating the local
real-quadrature noise variance to the one-sided PSD. Equivalently, the same known-phase
result may be written as $\delta A=\sqrt{2S_{yy}^{\mathrm{sym}}(\omega_m)B_w}$ in
double-sided symmetrized notation. Therefore
\begin{equation}
\delta A\simeq \sqrt{S_{yy}(\omega_m)B_w}.
\end{equation}
For a rectangular window one has
\begin{equation}
\delta A=\sqrt{\frac{S_{yy}}{T_{\mathrm{int}}}},
\end{equation}
whereas for a Hann window
\begin{equation}
\delta A=\sqrt{\frac{3S_{yy}}{2T_{\mathrm{int}}}}.
\end{equation}
If the phase is unknown and must be estimated from two orthogonal quadratures, the variance
doubles and $\delta A$ acquires the usual factor $\sqrt{2}$. If the signal frequency is not
exactly tracked, a leakage factor $\mathcal L\le 1$ should be inserted as
$A\rightarrow \mathcal L A$.

\section{Generic two-channel variational readout}
\label{app:variational}
The main text does not require any specific readout architecture beyond Eq.~(\ref{eq:readout}),
but we record the generic form of a two-channel estimator because future
semiclassical-gravity experiments may need such an upgrade. The phrase ``variational readout''
means that the experiment does not use the raw angle channel alone. It also measures an auxiliary
quadrature or auxiliary channel that carries information about the noise that will otherwise
appear as backaction. A frequency-dependent linear combination can then cancel part of the
correlated noise without canceling the signal.

Suppose two angle-equivalent channels are available,
\begin{align}
y_1(\omega) &= \theta(\omega)+n_1(\omega), \\
y_2(\omega) &= c_s(\omega)\theta(\omega)+c_b(\omega)\tau_{\ba}(\omega)+n_2(\omega),
\end{align}
where $n_1$ and $n_2$ are the intrinsic noises of the two readout channels. The coefficient
$c_s(\omega)$ is the calibrated signal transfer from the true angle to the auxiliary channel, and
$c_b(\omega)$ is the calibrated transfer from the backaction torque to that channel; their units
are whatever is required to make $y_2$ angle equivalent. The second channel could be realized, for
example, by a speed-meter-like optical response, an auxiliary cavity monitoring radiation-pressure
correlations, or a stroboscopic quadrature measurement. The cross spectra below are defined by
$S_{ij}(\omega)\equiv S_{N_iN_j}(\omega)$ for the noise parts $N_i$ of the two channels after the
known signal response has been removed. Consider the estimator
\begin{equation}
\theta_{\est}(\omega)=y_1(\omega)+\alpha(\omega)y_2(\omega).
\end{equation}
Its noise PSD is
\begin{equation}
S_{\est\est}(\omega)=S_{11}(\omega)+|\alpha(\omega)|^2S_{22}(\omega)
+2\,\mathrm{Re}\!\left[\alpha(\omega)S_{12}(\omega)\right].
\end{equation}
Minimizing over $\alpha$ gives the closed-form weight
\begin{equation}
\alpha_{\opt}(\omega)=-\frac{S_{12}(\omega)}{S_{22}(\omega)},
\end{equation}
and the corresponding minimum
\begin{equation}
S_{\est\est}^{\mathrm{min}}(\omega)=S_{11}(\omega)-\frac{|S_{12}(\omega)|^2}{S_{22}(\omega)}.
\end{equation}
This is the torsion-balance counterpart of standard variational readout in interferometric
optomechanics. The crucial point is that the auxiliary channel must be \emph{physically specified}
and calibrated through $c_s$, $c_b$, and the cross spectrum $S_{12}$; a formal reference to the
canonical momentum operator alone is not sufficient for an experimental proposal. In a concrete
optical realization $c_s$ and $c_b$ are generally not independent free functions: both are linked
by the same cavity or transducer dynamics, and the role of the generic treatment here is only to
state the optimal linear estimator once those coupled transfer functions have been calibrated.
In practice, $S_{12}$ should be measured with the same windowing and leakage treatment used in the
science analysis, ideally using coherent calibration lines and optical-power sweeps that isolate
the backaction pathway from purely electronic correlations.

\section{Minimal publication recipe for reusable residual spectra}
\label{app:publication}
For future torsion-balance papers intended to support semiclassical-gravity reinterpretation, the
minimum data product required for reuse is a residual equivalent-torque spectrum together with the
metadata needed to reconstruct it. This appendix is a reporting checklist, not an additional
physical assumption: it states what has to be published so that another group can tell which part
of an observed spectrum is thermal, Newtonian, readout, or still-unexplained residual noise. The
checklist is:
\begin{enumerate}
\item identify $I$, $\kappa$, and $Q$ from ringdown and/or driven transfer-function measurements;
\item state the PSD convention (one-sided or double-sided), the window, the ENBW, and any leakage
correction $\mathcal L$ used in the estimator;
\item provide the suspension-loss calibration, including the frequency band over which a constant
loss angle $\phi$ is a valid approximation;
\item publish the dominant auxiliary channels $x_j$ and their torque calibrations
$K_j=\partial\tau/\partial x_j$ so that Eq.~(\ref{eq:residual_recipe}) can be reconstructed;
\item when a fully subtracted residual is unavailable, state explicitly whether the quoted bound is
based on a residual spectrum or on the observed total floor.
\end{enumerate}
If a white-noise summary $D_\tau$ is quoted, it is also appropriate to state the leading calibration
uncertainty. A simple first-order estimate is
\begin{equation}
\frac{\delta D_\tau}{D_\tau}
\simeq
\left[
\left(
2\left\langle\frac{\delta|\chi|}{|\chi|}\right\rangle_{\mathcal B}
\right)^2
+
\left(
\left\langle\frac{\delta S_{yy}^{\mathrm{res},+}}{S_{yy}^{\mathrm{res},+}}\right\rangle_{\mathcal B}
\right)^2
\right]^{1/2},
\end{equation}
where the band average follows the same weighting used in Eq.~(\ref{eq:Dtau_band}).
This estimate assumes that the susceptibility calibration and the residual-spectrum estimate are
effectively independent; if the same suspension-model parameters enter both, their covariance
should be propagated explicitly rather than neglected.
This checklist records the minimum metadata needed to make the quantities in
Secs.~\ref{sec:deterministic} and \ref{sec:diffusion} portable across platforms.


\begin{thebibliography}{99}
\bibitem{GundlachMerkowitz2000}
J.~H.~Gundlach and S.~M.~Merkowitz,
``Measurement of Newton's constant using a torsion balance with angular acceleration feedback,''
Phys.\ Rev.\ Lett.\ \textbf{85}, 2869 (2000).

\bibitem{Schlamminger2006}
S.~Schlamminger, E.~Holzschuh, W.~K\"undig, F.~Nolting, R.~Pixley, and J.~Schurr,
``A measurement of Newton's gravitational constant,''
Phys.\ Rev.\ D \textbf{74}, 082001 (2006).

\bibitem{Quinn2013}
T.~J.~Quinn, H.~C.~Speake, S.~J.~Parks, and R.~S.~Davis,
``Improved determination of $G$ using two methods,''
Phys.\ Rev.\ Lett.\ \textbf{111}, 101102 (2013).

\bibitem{Wagner2012WEP}
T.~A.~Wagner, S.~Schlamminger, J.~H.~Gundlach, and E.~G.~Adelberger,
``Torsion-balance tests of the weak equivalence principle,''
Class.\ Quantum Grav.\ \textbf{29}, 184002 (2012).

\bibitem{Kapner2007}
D.~J.~Kapner, T.~S.~Cook, E.~G.~Adelberger, J.~H.~Gundlach,
B.~R.~Heckel, C.~D.~Hoyle, and H.~E.~Swanson,
``Tests of the gravitational inverse-square law below the dark-energy length scale,''
Phys.\ Rev.\ Lett.\ \textbf{98}, 021101 (2007).

\bibitem{Tan2020}
W.-H.~Tan \emph{et al.},
``Improvement for testing the gravitational inverse-square law at the submillimeter range,''
Phys.\ Rev.\ Lett.\ \textbf{124}, 051301 (2020).

\bibitem{PageGeilker1981}
D.~N.~Page and C.~D.~Geilker,
``Indirect evidence for quantum gravity,''
Phys.\ Rev.\ Lett.\ \textbf{47}, 979 (1981).

\bibitem{Yan2025}
T.~Yan, Y.~Liu, L.~Prokhorov, J.~Smetana, H.~Miao, Y.~Ma,
V.~Boyer, and D.~Martynov,
``First result for testing semiclassical gravity effect with a torsion balance,''
Phys.\ Rev.\ D \textbf{111}, 082007 (2025).

\bibitem{BoseRMP2025}
S.~Bose, A.~Mazumdar, G.~W.~Morley, H.~Ulbricht, M.~Toros,
M.~Paternostro, A.~A.~Geraci, P.~Barker, M.~S.~Kim, and G.~Milburn,
``Massive quantum systems as interfaces of quantum mechanics and gravity,''
Rev.\ Mod.\ Phys.\ \textbf{97}, 015003 (2025).

\bibitem{Caves1981}
C.~M.~Caves,
``Quantum-mechanical noise in an interferometer,''
Phys.\ Rev.\ D \textbf{23}, 1693 (1981).

\bibitem{ClerkRMP2010}
A.~A.~Clerk, M.~H.~Devoret, S.~M.~Girvin, F.~Marquardt, and R.~J.~Schoelkopf,
``Introduction to quantum noise, measurement, and amplification,''
Rev.\ Mod.\ Phys.\ \textbf{82}, 1155 (2010).

\bibitem{AspelmeyerRMP2014}
M.~Aspelmeyer, T.~J.~Kippenberg, and F.~Marquardt,
``Cavity optomechanics,''
Rev.\ Mod.\ Phys.\ \textbf{86}, 1391 (2014).

\bibitem{BraginskyKhaliliBook}
V.~B.~Braginsky and F.~Ya.~Khalili,
\emph{Quantum Measurement}
(Cambridge University Press, Cambridge, 1992).

\bibitem{Kimble2001}
H.~J.~Kimble, Y.~Levin, A.~B.~Matsko, K.~S.~Thorne, and S.~P.~Vyatchanin,
``Conversion of conventional gravitational-wave interferometers into quantum
nondemolition interferometers by modifying their input and/or output optics,''
Phys.\ Rev.\ D \textbf{65}, 022002 (2001).

\bibitem{MohantaShikano2026ANN}
J.~Mohanta and Y.~Shikano,
``Atmospheric Newtonian noise in torsion-balance determinations of the gravitational constant $G$,''
preprint (2026).

\bibitem{Giulini2022}
D.~Giulini, A.~Gro{\ss}ardt, and P.~K.~Schwartz,
``Coupling quantum matter and gravity,''
in \emph{Modified and Quantum Gravity: From Theory to Experimental Searches on All Scales},
edited by C.~Pfeifer and C.~L\"ammerzahl, Lecture Notes in Physics Vol.~1017
(Springer, Cham, 2023), pp.~491--550.

\bibitem{Oppenheim2023}
J.~Oppenheim,
``A postquantum theory of classical gravity?''
Phys.\ Rev.\ X \textbf{13}, 041040 (2023).

\bibitem{Layton2023}
I.~Layton, J.~Oppenheim, A.~Russo, and Z.~Weller-Davies,
``The weak field limit of quantum matter back-reacting on classical spacetime,''
JHEP \textbf{08}, 109 (2023).

\bibitem{FedidaKent2025}
S.~Fedida and A.~Kent,
``Mixture equivalence principles and postquantum theories of gravity,''
Phys.\ Rev.\ D \textbf{111}, 126016 (2025).

\bibitem{BassiRMP2013}
A.~Bassi, K.~Lochan, S.~Satin, T.~P.~Singh, and H.~Ulbricht,
``Models of wave-function collapse, underlying theories, and experimental tests,''
Rev.\ Mod.\ Phys.\ \textbf{85}, 471 (2013).

\bibitem{Bassi2017}
A.~Bassi, A.~Gro{\ss}ardt, and H.~Ulbricht,
``Gravitational decoherence,''
Class.\ Quantum Grav.\ \textbf{34}, 193002 (2017).

\bibitem{Carlesso2019}
M.~Carlesso, A.~Bassi, M.~Paternostro, and H.~Ulbricht,
``Testing the gravitational field generated by a quantum superposition,''
New J.\ Phys.\ \textbf{21}, 093052 (2019).

\bibitem{Grossardt2021}
A.~Gro{\ss}ardt,
``Dephasing and inhibition of spin interference from semi-classical self-gravitation,''
Class.\ Quantum Grav.\ \textbf{38}, 245009 (2021).

\bibitem{Altamura2024}
D.~G.~A.~Altamura, M.~Carlesso, S.~Donadi, and A.~Bassi,
``Noninterferometric rotational test of the continuous spontaneous localization model:
Enhancement of the collapse noise through shape optimization,''
Phys.\ Rev.\ A \textbf{109}, 062212 (2024).

\bibitem{Altamura2025}
D.~G.~A.~Altamura, A.~Vinante, and M.~Carlesso,
``Improved bounds on collapse models from rotational noise of the Laser Interferometer Space Antenna Pathfinder mission,''
Phys.\ Rev.\ A \textbf{111}, L020203 (2025).

\bibitem{Saulson1990}
P.~R.~Saulson,
``Thermal noise in mechanical experiments,''
Phys.\ Rev.\ D \textbf{42}, 2437 (1990).

\bibitem{Harms2015}
J.~Harms,
``Terrestrial gravity fluctuations,''
Living Rev.\ Relativ.\ \textbf{18}, 3 (2015).

\bibitem{Prokhorov2024}
J.~Heinze, L.~Prokhorov, J.~Smetana, \emph{et al.},
``Design and sensitivity of a 6-axis seismometer for gravitational wave detection and geoscience,''
Phys.\ Rev.\ D \textbf{109}, 042007 (2024).
\end{thebibliography}
\end{document}